\documentclass[12pt]{article}
\usepackage{graphicx}
\usepackage[utf8]{inputenc}
\usepackage[margin=1in]{geometry}
\usepackage[onehalfspacing]{setspace}
\usepackage{array}
\usepackage{changepage}
\usepackage{threeparttable}
\usepackage{multirow}
\usepackage{pdflscape}

\usepackage{amssymb}
\usepackage{amsmath}
\usepackage{amsthm}
\newtheorem{assumption}{Assumption}
\newtheorem{subassumption}{Assumption\normalfont}[assumption]

\newenvironment{assumption*}
{\ifnum\value{subassumption}=0 \stepcounter{assumption}\fi\subassumption}
{\endsubassumption}
\newenvironment{assumption+}[1]
{\renewcommand{\thesubassumption}{#1}\subassumption}
{\endsubassumption}

\renewcommand\footnotemark{}

\theoremstyle{definition}

\usepackage{todonotes}
\usepackage[english]{babel}
\usepackage{bbm,graphicx,bm}
\usepackage[colorlinks = TRUE, allcolors = black]{hyperref}
\usepackage{float} 
\usepackage{caption}
\usepackage[labelformat=simple,labelfont={}]{subcaption}

\usepackage{booktabs}
\usepackage{graphicx}
\usepackage{titlesec}

\newcommand{\E}{\ensuremath{\mathbb{E}}}
\newcommand{\Prob}{\ensuremath{\mathbb{P}}}

\newcommand{\Var}{\text{Var}}

\usepackage[normalem]{ulem}

\usepackage[authoryear,round]{natbib}

\titleformat{\section}
  {\normalfont\large\bfseries}
  {\thesection}{1em}{}

\titleformat{\subsection}
  {\normalfont\normalsize\bfseries}
  {\thesubsection}{1em}{}

\RequirePackage{xcolor}
\newif\ifshowchanges
\showchangestrue
\ifdefined\NoChanges\showchangesfalse\fi

\begin{document}

\onehalfspacing

\title{\textsc{\Large{A Practical Guide to Instrumental Variables Methods with Heterogeneous Treatment Effects}}\thanks{Tymon S{\l}oczy\'nski is Associate Professor of Economics, Brandeis University, Waltham, Massachusetts. Liyang Sun is Associate Professor of Economics, University College London, London, United Kingdom. S.~Derya Uysal is Associate Professor of Economics, LMU Munich, Munich, Germany. Their email addresses are tslocz@brandeis.edu, liyang.sun@ucl.ac.uk, and derya.uysal@econ.lmu.de.}}

\author{\textsc{Tymon S{\l}oczy\'nski}
\and
\textsc{Liyang Sun}
\and
\textsc{S. Derya Uysal}
}

\date{\textsc{November 9, 2026}}

\maketitle

\vspace{-0.275cm}

\noindent
Instrumental variables methods are widely used in economics. Roughly 30 percent of National Bureau of Economic Research (NBER) working papers in applied microeconomics mention instrumental variables \citep{GP2026}, and the NBER working papers also show that the  popularity of instrumental variables methods in applied microeconomics has grown continuously since the 1980s \citep{CKZ2020}.

Standard textbook treatments of instrumental variables methods assume a linear model with constant effects. However, constant effects may be implausible in many contexts. For example, consider a randomized controlled trial, in which the randomization provides a source of exogenous variation and the randomized assignment to treatment serves as the instrumental variable (``instrument'' for short). Participants may respond differently to treatment: for example, if the instrumental variable is a randomized offer to enroll in health insurance and the treatment is actual enrollment, some participants (say, older individuals) will react more strongly to insurance coverage than others. If the chance of being offered treatment also differs across groups (say, older vs.~younger individuals), researchers must adjust for these groups, and the key question becomes how to do so. The standard instrumental variables approach overweights groups where the chance of being offered treatment is closer to 50-50.

In this paper, we will offer a practical guide to the literature on instrumental variables methods with heterogeneous treatment effects, which originates from the work of Joshua Angrist and Guido Imbens on the so-called ``local average treatment effect'' (``LATE'') framework. The Nobel committee cited this work in awarding the 2021 Nobel Prize to Angrist and Imbens (shared also with David Card). The LATE framework, developed in \cite{IA1994}, \cite{AI1995}, and \cite{AIR1996}, allows for a very general form of treatment effect heterogeneity, which the textbook model rules out. This framework has become highly influential in applied microeconomics, but lessons from its recent extensions are less well-known. We highlight several areas in which modern applied work deviates from best practices motivated by the recent theoretical literature.

The first area we explore is the choice of the target parameter. When the instrument is valid only conditional on covariates and treatment effects differ across groups, common approaches to adjusting for covariates estimate different weighted averages of the group-specific effects, and not all of them equal the local average treatment effect. We discuss three approaches, including a preferred strategy to estimate the local average treatment effect when covariates matter. Second, we consider the flexibility of parametric specifications. With heterogeneous effects, the standard instrumental variables estimand has a causal interpretation only if the chance of being offered treatment is linear in the covariates. What happens if it is not? As one approach to such problems, we will discuss flexible estimation strategies based on machine learning. Third, we explore  possible violations of assumptions underlying the LATE framework, and in particular the often controversial assumption of monotonicity. These assumptions have testable implications that have spurred a sizable theoretical literature, yet have had limited influence on applied work. 

Throughout the paper, we focus primarily on an instrumental variables framework with a binary treatment and a binary instrument, like in the randomized controlled trial setting described above. However, we also offer some thoughts about more complex settings.

\section*{Three Target Parameters}

In this section, we compare three approaches to controlling for covariates and illustrate how the different parameters they estimate can lead to different substantive conclusions.

\subsection*{Motivating Example: Stratified Randomized Controlled Trials with Imperfect Compliance}

We start by considering a leading application of instrumental variables methods: a randomized controlled trial with imperfect compliance (``noncompliance'' for short). We assume that there is a binary treatment variable and a randomized assignment to treatment. We expect the randomized assignment to induce some individuals to receive treatment, but we also expect compliance to be imperfect: some individuals assigned to treatment may choose not to receive it and some individuals not assigned to receive it may end up doing so. In the end, the randomized assignment serves as an instrument, allowing us under certain assumptions to estimate the average causal effect of the treatment on the outcome for a particular group of individuals. Specifically, the independence assumption requires the instrument to be as good as randomly assigned, so that it is unrelated to how individuals would respond to being offered treatment and to actually being treated. The ``exclusion restriction'' requires the instrument to affect the outcome only indirectly, through the treatment, and otherwise to have no direct effect on it. The relevance assumption requires the instrument to change treatment take-up for at least some individuals. The monotonicity assumption requires the instrument to shift the treatment in the same direction for everyone who reacts to it, thus ruling out ``defiers'' who react to the instrument by moving in the opposite direction. We will return to a discussion of these assumptions, and the monotonicity assumption in particular, later in this paper. For now, we note that when the instrument is a randomized assignment to treatment, then independence holds by design, relevance is expected (and testable), while exclusion and monotonicity are often plausible. When these assumptions hold, the usual instrumental variables regression identifies the local average treatment effect, or LATE\@: the average treatment effect for ``compliers,'' who comply with their treatment assignment, taking the treatment if offered but not otherwise \citep{IA1994}. 

Importantly, however, treatment assignment is rarely \emph{completely} random. In a ``stratified'' randomized controlled trial, the researcher first divides the study population according to certain characteristics, and then carries out random assignment within each subgroup. For example,  the Oregon Health Insurance Experiment arose when the state of Oregon did not have sufficient funding to expand Medicaid coverage to all low-income adults. Thus, the state held a lottery in early 2008 for who would become eligible to apply for Medicaid coverage, in this way setting up a treatment and a control group for researchers to study the effect of Medicaid coverage on a range of health care usage, health, and financial outcomes.

In this experiment, Medicaid eligibility was, for practical purposes, randomly assigned within household-size strata. However, because of the design of the survey that collected outcome data, assignment probabilities in the survey sample vary across cells defined by household size \emph{and} timing of survey wave \citep{Finkelsteinetal2012}. The independence assumption no longer holds unconditionally, because these covariates may also be related to potential outcomes and treatments. Specifically, household size is plausibly related to health expenditures, health care utilization, and baseline health status, while survey wave may pick up time-varying shocks to outcomes. We need to control for these particular covariates.

\begin{table}[!t]
\begin{adjustwidth}{-1in}{-1in}
\centering
\begin{threeparttable}
\caption{Weights on Covariate-Specific LATEs}
\label{tab:LATE-weights}
\begin{tabular}{>{\centering\arraybackslash}m{1.5cm} >{\centering\arraybackslash}m{9.75cm} >{\centering\arraybackslash}m{4cm} >{\centering\arraybackslash}m{4.25cm}}
\hline\hline
Estimand & Estimation Procedure & $\omega_{m,j}$ & What's Overweighted? \\
\hline
$\beta_{\mathrm{IV}}$ & Standard linear specification, with $Z_i$ as the instrument for $D_i$ & $p_j\pi_j \cdot \Var \left[ Z_{i} \mid X_{ij} = 1 \right]$ & Cells where assignment varies more \\ \addlinespace[8pt]
$\beta_{\mathrm{AI}}$ & Standard linear specification, with interactions of $X_i$ and $Z_i$ as instruments for $D_i$ & $p_j\pi_j^2 \cdot \Var \left[ Z_{i} \mid X_{ij} = 1 \right]$ & Cells where the first stage is stronger and assignment varies more \\ \addlinespace[8pt]
$\beta_{\mathrm{LATE}}$ & LATE specification, with $Z_i$ as the instrument for $D_i$; or the average effect of $Z_i$ on $Y_i$ divided by the average effect of $Z_i$ on $D_i$ \citep{Frolich2007} & $p_j\pi_j$ & Nothing; cells are weighted in proportion to their complier shares \\
\hline
\end{tabular}
\begin{footnotesize}
\begin{tablenotes}[flushleft]
\item \textit{Notes:} $\omega_{m,j}$ denotes the (unnormalized) weight in the expression $\beta_m = \sum_j \frac{ \omega_{m,j}}{\sum_k \omega_{m,k}}\tau_j$ for estimand $m \in \left\lbrace \text{IV}, \text{AI}, \text{LATE} \right\rbrace$. $p_{j} = \Prob( X_{ij} = 1 )$ denotes the share of individuals with $X_{ij} = 1$. $\pi_j = \E[ D_{i}(1)-D_{i}(0) \mid X_{ij} = 1 ]$ denotes the conditional first stage, equal to the conditional proportion of compliers under the monotonicity assumption.
\end{tablenotes}
\end{footnotesize}
\end{threeparttable}
\end{adjustwidth}
\end{table}

There is not, however, a unique method to control for covariates. We will discuss three strategies to control for covariates in stratified randomized controlled trials with noncompliance, which we also summarize in Table \ref{tab:LATE-weights}. Of these, we recommend the third, which targets the average effect for all compliers; in this setting, it can be implemented as a simple instrumental variables regression.  The first two are variations of the classical instrumental variables method that estimates the standard linear specification where covariates are controlled in an additively separable fashion. 

The first approach is the simplest instrumental variables regression based on the standard linear specification. Let $Y_{i}$ be the outcome, $D_{i}$ the treatment, $Z_{i}$ the instrument, and $X_{i}$ the set of additional covariates. Then the specification can be written as:
\begin{align}
Y_{i} = \gamma'X_{i} + \beta D_{i} + e_{i}, \tag{standard linear specification}\label{eq:linear IV structural}
\end{align}
where $Z_i$ is used as the instrument for $D_i$. This results in the first stage that regresses the treatment variable on the covariates $X_{i}$ and the instrument $Z_{i}$, such that $D_{i} = \delta'X_{i} + \pi Z_{i} + v_{i}$. The predicted values of $D_{i}$ from this first-stage regression are then plugged into the outcome model above, which yields a population coefficient $\beta_{\text{IV}}$ on $D_{i}$\@. In the case of the Oregon Health Insurance Experiment, $Y_i$ is an outcome variable such as the number of outpatient visits, $D_i$ indicates Medicaid coverage, $Z_i$ indicates being selected in the lottery, and the covariates consist of indicators for different combinations of values of household size and survey wave. This means that the data are effectively divided into household size $\times$ survey wave cells, and every observation belongs to exactly one cell. When this condition holds, we say that the covariates ``saturate'' the model. To simplify the comparison with other approaches, we will assume for now that the covariates have this property. The method of analysis in \cite{Finkelsteinetal2012} is equivalent to estimating $\beta_{\text{IV}}$ using a saturated specification with household size $\times$ survey wave indicators.

The second approach uses the same standard linear specification, but modifies the first stage by interacting the covariates with the instrument: $D_{i} = \delta'X_{i}+ \sum_j \pi_j Z_{i} X_{ij}+v_{i}$. \cite{AP2009} refer to this specification as the ``saturate and weight'' approach. Following \cite{Sloczynski2026}, who calls it ``AI's specification'' after \cite{AI1995}, we refer to the population coefficient on $D_{i}$ as $\beta_{\text{AI}}$. With saturated covariates, the first-stage regression underlying $\beta_{\text{AI}}$ is correctly specified by design, and each coefficient on the interaction terms, $\pi_j$, may be interpreted as the causal effect of $Z_i$ on $D_i$ in cell $j$, which is the same as the proportion of compliers in that cell.

What we refer to as AI's specification features a heterogeneous first stage, but retains the additively separable outcome equation from the standard linear specification. To introduce the third approach, which we call the \emph{local average treatment effect (LATE) specification}, it is convenient to omit one element of the saturated vector $X_i$ and include an intercept. Also, let $\bar X=\E[X_i]$ denote the vector of covariate-cell proportions, so that $X_i - \bar {X}$ is the vector of demeaned cell indicators. The LATE specification augments the standard linear specification with interactions between the instrument and these demeaned covariates:
\begin{align}
Y_i \, = \, \alpha \, + \, \beta_{\mathrm{LATE}} \cdot D_i \, + \, \gamma' X_i \, + \, \eta' \left[ Z_i \cdot \left( X_i - \bar X \right) \right] \, + \, e_i. \tag{LATE specification}\label{eq:LATE}
\end{align}
This specification accommodates heterogeneity in both the first stage and the cell-specific LATEs, and enables us to target the overall average effect for compliers. Notice that it only includes a single endogenous explanatory variable, $D_i$. We also use a single instrument, $Z_i$.

Why does this method work? The simplest intuition combines two important theoretical results. The first is a textbook property of instrumental variables regressions with a single treatment and a single instrument: the effect of the treatment on the outcome is equal to the ratio of the reduced-form and first-stage coefficients on the instrument. In the LATE specification, the reduced form and the first stage amount to regressing the outcome and the treatment, respectively, on a constant, $Z_i$, $X_i$, and the interactions $Z_i \cdot \left( X_i - \bar X \right)$.\footnote{The resulting first stage gives the same fitted values as in AI's specification. To see this, note that the first stage in AI's specification can be reparameterized as $D_i = \delta + \pi Z_i + \kappa' X_i + \phi' \left[ Z_i \cdot \left( X_i - \bar X \right) \right] + v_i$ without changing its fitted values. The LATE specification is also briefly discussed by \cite{ZDL2026}, who contrast it with their ``interacted two-stage least squares'' specification, which instead includes interactions between the treatment and appropriately centered covariates in the outcome equation.} Because the full set of interactions is included in both regressions and the interacted covariates are demeaned, the reduced-form and first-stage coefficients on $Z_i$ capture the average effects of $Z_i$ on the outcome and treatment. The second relevant result is the nonparametric identification result in \cite{Frolich2007}, according to which the local average treatment effect is identified \emph{exactly} as the ratio above: the average effect of the instrument on the outcome divided by the average effect of the instrument on the treatment. This result relies on the LATE assumptions extended to settings with covariates \citep[cf.][]{Abadie2003}, but does not impose constant treatment effects. The two theoretical results jointly imply that the coefficient on $D_i$ in the LATE specification recovers the overall average effect for compliers.

\citeauthor{Frolich2007}'s nonparametric identification result holds even when the set of covariates $X_i$ contains general, rather than discrete and saturated, covariates. Thus, the LATE specification can also be used when the covariates are not saturated. In that case, however, it only provides a linear approximation to the conditional means of the outcome and the treatment that enter the numerator and denominator. Alternatively, one may instead use ``inverse probability weighting'' or ``doubly robust'' methods to estimate the numerator and denominator in \citeauthor{Frolich2007}'s formula. With saturated discrete covariates, standard implementations of these approaches are numerically equivalent and reduce to computing the effect of the instrument on the outcome and treatment within each covariate cell, using differences in means, and then aggregating the cell-specific effects using the proportions of observations in the respective cells. When $X_i$ is not saturated, however, these approaches need not coincide because they rely on different approximations for the relevant conditional means and instrument propensity scores, a point to which we return below.

The key point here is that, when treatment effects are heterogeneous, $\beta_{\text{IV}}$, $\beta_{\text{AI}}$, and $\beta_{\text{LATE}}$ are generally not the same; each parameter corresponds to a different object with its specific interpretation. To see this, under the same assumptions as above (and with saturated covariates for simplicity), we can derive a weighted average representation of the three estimands in terms of covariate-specific local average treatment effects, $\tau_j$, that is, the average treatment effects for compliers in cell $j$. Indeed, each estimand can be written as $\beta_m = \sum_j w_{m,j} \cdot \tau_j$, where the weights $w_{m,j}$ are normalized versions of the unnormalized weights, $\omega_{m,j}$, in Table \ref{tab:LATE-weights}. The representations of $\beta_{\text{IV}}$ and $\beta_{\text{AI}}$, given in Table \ref{tab:LATE-weights}, follow from \cite{Sloczynski2026}; the representation of $\beta_{\mathrm{AI}}$ is also implied by Theorem~3 of \cite{AI1995}. The representation of $\beta_{\mathrm{LATE}}$ follows from its definition.

The expression for $\beta_{\mathrm{LATE}}$ is the most intuitive, aggregating $\tau_j$ based on the probability of compliance at each covariate level. In other words, each (normalized) weight is the proportion of all compliers in the corresponding cell. This means that $\beta_{\mathrm{LATE}}$ is equal to the overall average treatment effect for compliers. On the other hand, $\beta_{\text{IV}}$ and $\beta_{\text{AI}}$ introduce additional terms to their respective weights, which complicates interpretation. Specifically, $\beta_{\mathrm{IV}}$ overweights the cells where assignment varies more, that is, with large values of $\Var \left[ Z_{i} \mid X_{ij} = 1 \right]$. Similarly, $\beta_{\mathrm{AI}}$ overweights the cells where the first stage is stronger and the assignment varies more, according to the product of $\pi_j$ and $\Var \left[ Z_{i} \mid X_{ij} = 1 \right]$ \citep{Sloczynski2026}. In a model with homogeneous treatment effects in which the standard linear specification is appropriate, such weighting may be desirable because it improves estimation efficiency; with heterogeneous treatment effects, however, it would be difficult to justify an explicit interest in $\beta_{\mathrm{IV}}$ or $\beta_{\mathrm{AI}}$. Ultimately, the average treatment effect for compliers, $\beta_{\text{LATE}}$, may differ substantially from $\beta_{\mathrm{IV}}$, $\beta_{\mathrm{AI}}$, or both. Such differences can arise when $\pi_j$, the conditional proportion of compliers, or $\Var \left[ Z_{i} \mid X_{ij} = 1 \right]$, the conditional variance of the instrument, vary across cells.\footnote{The interpretation of these parameters will change when the monotonicity assumption is violated. We return to this issue below.} Whether these differences are large in a given context, however, is an empirical question.

\begin{table}[!t]
\begin{adjustwidth}{-1in}{-1in}
\centering
\begin{threeparttable}
\caption{Causal Effects of Medicaid on the Number of Outpatient Visits \normalsize{\label{tab:OHIE main}}}
\begin{tabular}{>{\centering\arraybackslash}m{2cm} >{\centering\arraybackslash}m{2cm} >{\centering\arraybackslash}m{3cm} >{\centering\arraybackslash}m{2cm} >{\centering\arraybackslash}m{2cm} >{\centering\arraybackslash}m{2cm} >{\centering\arraybackslash}m{2cm} >{\centering\arraybackslash}m{2cm}}
    \hline\hline
    & \multicolumn{3}{c}{A. Full Sample} & $\hat{\beta}_{\mathrm{LATE}}$ & $\hat{\beta}_{\mathrm{IV}}$ & $\hat{\beta}_{\mathrm{AI}}$ & \\
    \hline
    & & & & 1.001 & 1.001 & 1.010 & \\
    & & & & (0.134) & (0.134) & (0.132) & \\[5pt]
    \hline
    & \multicolumn{3}{c}{B. Three-Person Households} & $\hat{\beta}_{\mathrm{LATE}}$ & $\hat{\beta}_{\mathrm{IV}}$ & $\hat{\beta}_{\mathrm{AI}}$ & \\
    \hline
    & & & & 4.022 & 2.790 & 2.268 & \\
    & & & & (1.189) & (1.296) & (1.118) & \\[5pt]
    \hline
    \multicolumn{8}{c}{C. Estimated Weights in the Sample of Three-Person Households} \\
    \hline
    Survey Wave & $\hat{p}_j$ & $\widehat{\Var} \left[ Z_{i} \mid X_{ij} = 1 \right]$ & $\hat{\pi}_j$ & $\hat{w}_{\text{LATE},j}$ & $\hat{w}_{\text{IV},j}$ & $\hat{w}_{\text{AI},j}$ & $\hat{\tau}_j$ \\
    \hline
    1 & 0.241 & 0.168 & 0.455 & 0.351 & 0.600 & 0.723 & 1.067 \\
    2 & 0.448 & 0.037 & 0.280 & 0.401 & 0.151 & 0.112 & 6 \\
    3 & 0.310 & 0.099 & 0.250 & 0.248 & 0.249 & 0.165 & 5 \\
    \hline
\end{tabular}
\begin{footnotesize}
\begin{tablenotes}[flushleft]
\item \textit{Notes:} The source of data is \cite{Finkelsteinetal2012}. The outcome is the number of outpatient visits in the last six months, as in Table V of \cite{Finkelsteinetal2012}. The sample sizes are 23,441 and 58 in Panels A and B, respectively. Standard errors are in parentheses and are clustered at the household level. Panel C is also restricted to the subsample of 58 individuals in three-person households. $\hat{p}_j$ is the proportion of observations in covariate cell (survey wave) $j$. $\widehat{\Var} \left[ Z_{i} \mid X_{ij} = 1 \right]$ is the variance of $Z_{i}$ in cell $j$. $\hat{\pi}_j$ is the estimated proportion of compliers in cell $j$. $\hat{w}_{\text{LATE},j}$, $\hat{w}_{\text{IV},j}$, and $\hat{w}_{\text{AI},j}$ are the estimated (normalized) weights of cell $j$ in $\beta_{\text{LATE}}$, $\beta_{\mathrm{IV}}$, and $\beta_{\mathrm{AI}}$, respectively. $\hat{\tau}_j$ is the estimated LATE on the number of outpatient visits in cell $j$. Each of $\hat{\beta}_{\mathrm{LATE}}$, $\hat{\beta}_{\mathrm{IV}}$, and $\hat{\beta}_{\mathrm{AI}}$, as reported in Panel B, can be obtained as the dot product of $\hat{\tau}_j$ and the respective weights, i.e., $\hat{\beta}_m = \sum_j \hat{w}_{m,j} \cdot \hat{\tau}_j$ for $m \in \left\lbrace \text{LATE}, \text{IV}, \text{AI} \right\rbrace$.
\end{tablenotes}
\end{footnotesize}
\end{threeparttable}
\end{adjustwidth}
\end{table}

To illustrate this empirical question, we use the original survey data from the Oregon Health Insurance Experiment \citep{Finkelsteinetal2012}. In this setting, as mentioned above, the instrument (the ability to apply for Medicaid) is randomized conditional on household size and survey wave, and the identifying assumptions underlying the local average treatment effect framework seem plausible. For our illustration, we focus on the impact of Medicaid coverage on the number of outpatient visits in the last six months. To simplify the illustration, unlike \cite{Finkelsteinetal2012}, we do not use survey weights.

Panel A of Table \ref{tab:OHIE main} reports the estimates of $\beta_{\text{LATE}}$, $\beta_{\mathrm{IV}}$, and $\beta_{\mathrm{AI}}$ using the full sample from the Oregon Health Insurance Experiment\@. Although there is substantial heterogeneity in covariate-specific local average treatment effects, the different estimands yield nearly identical point estimates and standard errors, suggestive of a precisely estimated increase in the number of outpatient visits in response to Medicaid enrollment by about one in six months on average. Apparently, the correlation between $\widehat{\Var} \left[ Z_{i} \mid X_{ij} = 1 \right]$ and $\hat{\tau}_j$ as well as $\hat{\pi}_j \cdot \widehat{\Var} \left[ Z_{i} \mid X_{ij} = 1 \right]$ and $\hat{\tau}_j$ is sufficiently weak among compliers to render the differences between $\hat{\beta}_{\mathrm{LATE}}$, $\hat{\beta}_{\mathrm{IV}}$, and $\hat{\beta}_{\mathrm{AI}}$ meaningless.

The picture is very different, however, when we restrict our attention to the subsample of three-person households in Panel B of Table \ref{tab:OHIE main}. This restriction does not alter the identifying assumptions, as independence continues to hold conditional on survey wave, but it changes the practical importance of the different components in the weighted average representations in  Table \ref{tab:LATE-weights}. Indeed, the estimated $\beta_{\text{LATE}}$, 4.022, is now substantially larger than $\hat{\beta}_{\mathrm{IV}}$ and $\hat{\beta}_{\mathrm{AI}}$, which are equal to 2.790 and 2.268, respectively.

The differences between $\hat{\beta}_{\mathrm{LATE}}$, $\hat{\beta}_{\mathrm{IV}}$, and $\hat{\beta}_{\mathrm{AI}}$ follow directly from their distinct weighting schemes, summarized in Table \ref{tab:LATE-weights}. The estimated $\beta_{\text{LATE}}$ aggregates covariate-specific treatment effects in proportion to each cell's estimated share among compliers. By contrast, $\hat{\beta}_{\mathrm{IV}}$ also weights each cell by the conditional variance of the instrument, $\widehat{\Var} \left[ Z_{i} \mid X_{ij} = 1 \right]$, while $\hat{\beta}_{\mathrm{AI}}$ further weights each cell by $\hat{\pi}_j$. As a result, these estimates place relatively more weight on covariate cells with greater identifying variation, which need not be the cells that are most representative of the complier population.

Panel C of Table \ref{tab:OHIE main} illustrates this mechanism. For each cell in the subsample of three-person households, we report its sample proportion, $\hat{p}_j$, the conditional variance of $Z_{i}$, as well as the cell-specific estimates of the proportion of compliers, $\hat{\pi}_j$, and the local average treatment effect, $\hat{\tau}_j$. We also report the weights underlying each estimate, which are the normalized sample counterparts of the expressions in Table \ref{tab:LATE-weights}. For example, $\hat{w}_{\text{LATE},j} = \hat{p}_j \hat{\pi}_j / \sum_k \hat{p}_k \hat{\pi}_k$ and $\hat{w}_{\text{IV},j} = \hat{p}_j \hat{\pi}_j \cdot \widehat{\Var} \left[ Z_{i} \mid X_{ij} = 1 \right] / \sum_k \hat{p}_k \hat{\pi}_k \cdot \widehat{\Var} \left[ Z_{i} \mid X_{ik} = 1 \right]$. In the first covariate cell, $\hat{w}_{\text{LATE},1} = \left( 0.241 \cdot 0.455 \right) / \left( 0.241 \cdot 0.455 + 0.448 \cdot 0.280 + 0.310 \cdot 0.250 \right) \approx 0.351$ and $\hat{w}_{\text{IV},1} = \left( 0.241 \cdot 0.168 \cdot 0.455 \right) \; / \; \left( 0.241 \cdot 0.168 \cdot 0.455 + 0.448 \cdot 0.037 \cdot 0.280 + 0.310 \cdot 0.099 \cdot 0.250 \right) \approx 0.600$. (All values are rounded for illustration.) Each of the estimates reported in Panel B of Table \ref{tab:OHIE main} can be obtained as the dot product of $\hat{\tau}_j$ and the respective weights. For example, $\hat{\beta}_{\mathrm{LATE}} = 0.351 \cdot 1.067 + 0.401 \cdot 6 + 0.248 \cdot 5 \approx 4.022$ and $\hat{\beta}_{\mathrm{IV}} = 0.600 \cdot 1.067 + 0.151 \cdot 6 + 0.249 \cdot 5 \approx 2.790$. In this subsample, because $\widehat{\Var} \left[ Z_{i} \mid X_{ij} = 1 \right]$ and $\hat{\pi}_j$ are large when $\hat{\tau}_j$ is small, $\hat{\beta}_{\mathrm{IV}}$ and $\hat{\beta}_{\mathrm{AI}}$ underweight the covariate cells with large estimated treatment effects, resulting in estimates that are much smaller than $\hat{\beta}_{\mathrm{LATE}}$\@. The three underlying estimands answer different causal questions despite relying on the same experimental variation.

What are the specific causal questions answered by $\beta_{\text{LATE}}$, $\beta_{\mathrm{IV}}$, and $\beta_{\mathrm{AI}}$? In this application, $\beta_{\mathrm{LATE}}$ is the average effect of Medicaid coverage on outpatient visits for all compliers in three-person households, that is, for individuals who only enroll in Medicaid when they are selected in the lottery. On the other hand, $\beta_{\mathrm{IV}}$ and $\beta_{\mathrm{AI}}$ correspond to average effects of Medicaid coverage for \emph{data-driven subsets} of the complier population. For example, these subsets include disproportionately many first-wave compliers, who have the smallest estimated average treatment effect. Specifically, the estimates in Panel C of Table \ref{tab:OHIE main} suggest that the first-wave compliers correspond to 35 percent of all compliers, but they make up 60 and 72 percent, respectively, of the implicit complier subpopulations whose average treatment effects are recovered by $\beta_{\mathrm{IV}}$ and $\beta_{\mathrm{AI}}$\@. \cite{PS2026} show how to estimate the sizes of these implicit subpopulations and propose this measure as a simple diagnostic, with smaller subpopulation sizes reflecting a greater loss of internal validity relative to $\beta_{\mathrm{LATE}}$\@.\footnote{This diagnostic can be computed using the Stata package \texttt{causalrep}.} These implicit subpopulations are of no substantive interest: their composition reflects differences in assignment probabilities across waves, which follow from the survey design, and, for $\beta_{\mathrm{AI}}$, differences in the strength of the first stage.

Perhaps the most fundamental lesson of this essay is that a cautious researcher should investigate whether a similar pattern arises in their specific setting. Alternatively, it may be preferable to focus on $\beta_{\text{LATE}}$ as the main parameter of interest from the outset.

\subsection*{Beyond Stratified Randomized Controlled Trials}

In a stratified randomized controlled trial with noncompliance, cell indicators saturate the relevant covariate space by construction; that is, every observation belongs to exactly one cell. In other instrumental variables applications, however, saturated specifications may be infeasible, especially when covariates are numerous or continuously distributed. In such settings, interpreting the usual instrumental variables estimand, $\beta_{\mathrm{IV}}$, as a weighted average of causal effects requires an additional parametric restriction: the conditional mean of the instrument $Z_{i}$ given the covariates $X_{i}$ must be linear in $X_{i}$. \cite{BBMT2026} call this assumption ``rich covariates,'' and show that it is necessary for a causal interpretation of $\beta_{\mathrm{IV}}$.

On the other hand, when the covariate specification is not saturated, the nonparametric identification result in \cite{Frolich2007} continues to apply. As discussed above, this result implies that $\beta_{\mathrm{LATE}}$ can be estimated as the ratio of the average effect of the instrument on the outcome to its average effect on the treatment. Without saturation, these two average effects can be estimated in a number of distinct ways, implying a corresponding range of suitable estimators of $\beta_{\mathrm{LATE}}$\@.\footnote{Some of these estimators include those in \cite{Tan2006}, \cite{Frolich2007}, \cite{Uysal2011}, \cite{Heiler2022}, \cite{SSX2022}, \cite{SUW_DR,SUW_kappa}, \cite{MSSU2026}, and \cite{ZDL2026}. Related machine learning approaches have been developed by \cite{BCFH2017}, \cite{Chernozhukovetal2018}, \cite{ST2022}, \cite{SS2024}, and others.} In principle, one approach to estimating $\beta_{\mathrm{LATE}}$ is based on the LATE specification above. However, we might be uncomfortable with the implicit linearity assumptions that materialize in this specification in the absence of saturation. In what follows, we instead focus on an ``inverse probability weighting'' estimator of the same parameter, both to make the contrast with $\hat{\beta}_{\mathrm{IV}}$ clearer and to set up our additional discussion of the ``rich covariates'' assumption below. The inverse probability weighting estimator is based on a first-step estimate of the conditional mean of $Z_{i}$ given $X_{i}$, often called the instrument propensity score. Notice that this is the same conditional mean that is central to the ``rich covariates'' assumption. Here, however, the goal is to estimate $\beta_{\mathrm{LATE}}$. In practice, because $Z_{i}$ is binary, most researchers would typically use a logit or a probit model rather than the linear probability model implicit in $\beta_{\mathrm{IV}}$. The resulting instrument propensity score estimates are used to reweight observations with $Z_i=1$ and $Z_i=0$, that is, those offered treatment and those not offered it, so that the two subsamples are comparable in terms of their covariates. This approach enables straightforward estimation of the average effects of the instrument on the outcome and treatment using differences in reweighted means. The ratio of these two differences is the inverse probability weighting estimator proposed by \cite{Frolich2007} for the local average treatment effect.\footnote{\cite{SUW_kappa} emphasize the importance of using normalized weights in this context, explore the connection to the ``kappa weighting'' of \cite{Abadie2003}, and develop the Stata package \texttt{kappalate}. The package allows binary and nonbinary treatments, whether discrete or continuous; the instrument, however, must be binary.}

We continue to recommend focusing on estimation of $\beta_{\mathrm{LATE}}$ when covariates matter but saturated specifications are infeasible. The two parameters, $\beta_{\mathrm{IV}}$ and $\beta_{\mathrm{LATE}}$, may often be similar, but when they differ, $\beta_{\mathrm{LATE}}$ is arguably of much greater substantive interest. In what follows, because $\beta_{\mathrm{AI}}$ is also unlikely to be the main object of interest when it differs from $\beta_{\mathrm{LATE}}$, we focus on the comparison between $\beta_{\mathrm{IV}}$ and $\beta_{\mathrm{LATE}}$ for simplicity.

\section*{Parametric Misspecification}

In this section, we discuss how omitted interaction and higher-order terms can bias estimates of $\beta_{\mathrm{IV}}$ and $\beta_{\mathrm{LATE}}$, even when all requisite covariates are included. We describe tests for parametric misspecification and recommend machine learning as one way to address this problem.

\subsection*{Rich Covariates}

A substantial literature on weighted average representations of $\beta_{\mathrm{IV}}$, $\beta_{\mathrm{AI}}$, and related estimands has been based on the assumption that the conditional mean of the instrument $Z_{i}$ given the covariates $X_{i}$ is linear in $X_{i}$\@. For a binary instrument defined as assignment to treatment, this means that the probability of assignment is a linear function of the covariates. \cite{BBMT2026} show that this assumption, termed ``rich covariates,'' is not only sufficient but also \emph{necessary} for the resulting estimand to represent a nonnegatively weighted average of complier causal effects. If covariates are not rich, the estimand is not ``weakly causal,'' which means that when all conditional average treatment effects have the same sign, the estimand might even assume the opposite sign. This is equivalent to the absence of any subpopulation of compliers whose average effect will be recovered by the estimand of interest regardless of the pattern of treatment effect heterogeneity \citep{PS2026}.

An immediate implication is that applied researchers targeting simple instrumental variables estimands should only use covariate specifications that are plausibly rich. To examine whether a given specification is rich, \cite{BBMT2026} recommend the regression specification error test (RESET) of \cite{Ramsey1969}. In its canonical form, RESET, as applied to this problem, consists of estimating a regression of $Z_{i}$ on $X_{i}$ using ordinary least squares, obtaining fitted values, $\hat{Z}_{i}$, and, finally, estimating a regression of $Z_{i}$ on $X_{i}$ and several higher-order terms of $\hat{Z}_{i}$, such as $\hat{Z}_{i}^2$, $\hat{Z}_{i}^3$, and $\hat{Z}_{i}^4$ (again using ordinary least squares)\@. In this setup, RESET is the $F$ test of joint significance of the higher-order terms in the second-step regression. It should be interpreted as a test for neglected nonlinearity and should not be expected to detect any other omitted variable problems \citep[][Section 6.3.3]{Wooldridge2010}. If we reject the null hypothesis, we conclude that the original specification is not rich.

Estimators of $\beta_{\mathrm{LATE}}$ can suffer from similar specification problems. For example, the inverse probability weighting estimator discussed in the previous section relies on the assumption that the conditional mean of $Z_{i}$ given $X_{i}$ is correctly specified, \emph{however it is specified}. We can use, say, a logit or a probit model, which gives us more flexibility than the linear probability model implicit in $\beta_{\mathrm{IV}}$\@. Still, the logit or probit specification will not be correctly specified if we fail to include relevant interaction and higher-order terms, as in the case of ``rich covariates.'' To test for parametric misspecification of the logit or probit model, one may use the extension of RESET proposed by \cite{PW1996}. Here, we need to augment the original specification with several higher-order terms of the fitted linear index and, again, test for their joint significance. If we reject the null hypothesis, we conclude that the inverse probability weighting estimator of $\beta_{\mathrm{LATE}}$ will be inconsistent.

To illustrate the problem of parametric misspecification (and also to show that differences between alternative target parameters persist in a nonsaturated setting), we estimate $\beta_{\mathrm{IV}}$ and $\beta_{\mathrm{LATE}}$ using \citeauthor{Abadie2003}'s (\citeyear{Abadie2003}) sample from the 1991 Survey of Income and Program Participation (SIPP)\@. In this application, based on \cite{PVW1994}, one of the causal effects of interest is that of participation in a 401(k) retirement plan on participation in an individual retirement account (IRA), a question that bears on whether 401(k) saving substitutes for other forms of retirement saving. While 401(k) participation is likely endogenous, \cite{PVW1994} and \cite{Abadie2003} argue that 401(k) eligibility, which is determined by the employer, can be used as an instrument for 401(k) participation. Because 401(k) eligibility is not randomized or ``as good as randomly assigned,'' \cite{Abadie2003} controls for family income, age, age squared, marital status, and family size to estimate the causal effects of 401(k) participation using this instrument.

\begin{table}[!t]
\begin{adjustwidth}{-1in}{-1in}
\centering
\begin{threeparttable}
\caption{Causal Effects of 401(k) on IRA Participation \normalsize{\label{tab:401(k) reset}}}
\begin{tabular}{>{\centering\arraybackslash}m{7.5cm} >{\centering\arraybackslash}m{2cm} >{\centering\arraybackslash}m{2cm} >{\centering\arraybackslash}m{2cm}}
    \hline\hline
    \multirow{2}[0]{*}{A. Original Specification} & \multirow{2}[0]{*}{$\hat{\beta}_{\mathrm{IV}}$} & $\hat{\beta}_{\mathrm{LATE}}$ & $\hat{\beta}_{\mathrm{LATE}}$ \\
     & & (logit) & (probit) \\
    \hline
	 & 0.0274 & 0.0165 & 0.0175 \\
	 & (0.0132) & (0.0135) & (0.0136) \\
	 \\
	RESET $p$-Value & 0.000 & 0.000 & 0.000 \\
    \hline
    \multirow{2}[0]{*}{B. Flexible Specification} & \multirow{2}[0]{*}{$\hat{\beta}_{\mathrm{IV}}$} & $\hat{\beta}_{\mathrm{LATE}}$ & $\hat{\beta}_{\mathrm{LATE}}$ \\
     & & (logit) & (probit) \\
    \hline
	 & 0.0167 & 0.0177 & 0.0178 \\
	 & (0.0132) & (0.0128) & (0.0128) \\
	 \\
	RESET $p$-Value & 0.346 & 0.153 & 0.245 \\
	\hline
\end{tabular}
\begin{footnotesize}
\begin{tablenotes}[flushleft]
\item \textit{Notes:} The data are \citeauthor{Abadie2003}'s (\citeyear{Abadie2003}) subsample of the 1991 Survey of Income and Program Participation (SIPP)\@. The sample size is 9,275. The outcome is an indicator for participation in IRAs, as in Table 3 of \cite{Abadie2003}. The original covariate specification in Panel A includes family income, age, age squared, marital status, and family size. The flexible covariate specification in Panel B includes a quintic in family income, a quintic in age, an indicator for marital status, and indicators for each value of family size. $\hat{\beta}_{\mathrm{LATE}}$ corresponds to the normalized inverse probability weighting estimator of \cite{Uysal2011} and \cite{SUW_kappa}, with logit or probit instrument propensity scores estimated using maximum likelihood. RESET is based on a quartic in fitted values from each initial model of the instrument propensity score. Heteroskedasticity-robust standard errors are in parentheses. Conventional inference based on these standard errors assumes that 401(k) eligibility is a strong instrument. Unreported Anderson--Rubin confidence sets, which are robust to weak instruments, are similar to the conventional confidence intervals. For an overview of instrument strength and weak-instrument-robust inference in this journal, see Lee and Porter (2026).
\end{tablenotes}
\end{footnotesize}
\end{threeparttable}
\end{adjustwidth}
\end{table}

Panel A of Table \ref{tab:401(k) reset} replicates \citeauthor{Abadie2003}'s (\citeyear{Abadie2003}) estimate of $\beta_{\mathrm{IV}}$ based on this specification. This estimate, statistically significant at the 5 percent level, suggests that participation in a 401(k) retirement plan increases the probability of IRA participation by about 2.74 percentage points. However, Panel A of Table \ref{tab:401(k) reset} also reports two inverse probability weighting estimates of the local average treatment effect. These estimates, based on logit and probit instrument propensity scores, are much smaller than $\hat{\beta}_{\mathrm{IV}}$, with $p$-values equal to 0.221 and 0.197, respectively. This calls into question the conclusion about the positive impact of 401(k) on IRA participation. In our view, the possibility that such differences may materialize necessitates explicit estimation of $\beta_{\mathrm{LATE}}$ in relevant applied work---either as the main parameter of interest or at least as a robustness check.

With regard to the problem of parametric misspecification, Panel A of Table \ref{tab:401(k) reset} also reports the RESET $p$-values associated with each estimate. With the original covariate specification, we decidedly reject both the ``rich covariates'' assumption and  also the correct specification of the logit and probit models for the instrument propensity score. Can a more flexible specification salvage a causal interpretation of our estimates? Panel B of Table \ref{tab:401(k) reset} uses a quintic in family income, a quintic in age, an indicator for marital status, and indicators for each value of family size. With this set of covariates, RESET does not reject any of the null hypotheses of correct specification of the linear probability, logit, and probit models. $\hat{\beta}_{\mathrm{IV}}$ and $\hat{\beta}_{\mathrm{LATE}}$ are now very similar to each other, although $\hat{\beta}_{\mathrm{LATE}}$ has barely changed relative to Panel A while $\hat{\beta}_{\mathrm{IV}}$ has decreased by about 40 percent.

This illustration raises the question of whether $\hat{\beta}_{\mathrm{IV}}$ is less robust to parametric misspecification than estimators of $\beta_{\mathrm{LATE}}$. Although we are not aware of such results for generic inverse probability weighting estimators, other approaches to estimating $\beta_{\mathrm{LATE}}$ offer improved robustness properties, including the covariate balancing estimators of \cite{Heiler2022}, \cite{SSX2022}, and \cite{SUW_kappa}, and the doubly robust estimators of \cite{Tan2006}, \cite{Uysal2011}, \cite{SUW_DR}, and \cite{MSSU2026}.\footnote{Several of these estimators are implemented in the Stata package \texttt{drlate} \citep{SUW_DR}, which also accommodates models for count outcomes with many zeros, such as the number of outpatient visits in Table \ref{tab:OHIE main}.} We recommend wider use of these covariate balancing and doubly robust estimators, together with their machine learning counterparts discussed below.

\subsection*{Double Machine Learning}

The previous subsection explained that a causal interpretation of $\beta_{\mathrm{IV}}$ and $\beta_{\mathrm{LATE}}$ may fail when the conditional mean of the instrument given covariates suffers from parametric misspecification. One practical response is to manually enrich the covariate specification---that is, include additional interaction or higher-order terms---and then apply RESET\@. In many empirical applications, however, the number of potential interactions and nonlinear transformations is large, making manual specification and specification testing cumbersome.

A complementary approach is to use ``double machine learning'' to flexibly approximate the relevant conditional means. Double machine learning for estimating $\beta_{\mathrm{LATE}}$ has been studied by \cite{BCFH2017}, \cite{Chernozhukovetal2018}, \cite{ST2022}, and \cite{SS2024}, whereas \cite{Chernozhukovetal2018} also applied this methodology to $\beta_{\mathrm{IV}}$\@.

An application of double machine learning in an instrumental variables context proceeds as follows. The researcher supplies a rich dictionary of covariates to a machine learning algorithm, such as LASSO (``least absolute shrinkage and selection operator''), ridge, or regression tree. For an overview of these and other machine learning methods in this journal, useful starting points are \cite{varian2014big} and \cite{mullainathan2017machine}. Double machine learning then approximates the relevant nonlinear relationships---the conditional means of potential outcomes and potential treatments, as well as that of the instrument---in a data-driven way, with tuning parameters typically chosen by cross-validation. Finally, double machine learning  combines the resulting flexible predictions with orthogonal scores and cross-fitting to decrease the underlying ``regularization bias,'' a problem that arises when machine learning tools introduce bias in order to control variance, and in doing so may introduce systematic estimation errors. In this sense, double machine learning can be viewed as an automated approach to constructing ``rich'' covariate specifications, reducing the need for ad hoc choices of interaction and higher-order terms.

The outcome of using double machine learning will depend on the underlying choice of the machine learning algorithm. For example, LASSO assumes that the target functions can be well approximated by a sparse subset from the dictionary of covariates, whereas tree-based methods rely on recursive partitioning that builds up a decision tree by splitting the data into groups, followed by local averaging. Thus, even in a double machine learning framework, the validity of the researcher's conclusions continues to hinge on the chosen approximation. Accordingly, it is recommended to compare multiple machine learning algorithms to assess robustness and to reduce reliance on the approximation properties, and thus the implicit parametric restrictions, of any single method.

\begin{table}[!t]
\begin{adjustwidth}{-1in}{-1in}
\centering
\begin{threeparttable}
\caption{Double Machine Learning Estimates of Causal Effects of 401(k) on IRA Participation \normalsize{\label{tab:401(k) dml}}}
\begin{tabular}{>{\centering\arraybackslash}m{7.5cm} >{\centering\arraybackslash}m{2.75cm} >{\centering\arraybackslash}m{2.75cm}}
    \hline\hline
    & $\hat{\beta}_{\mathrm{IV}}$ & $\hat{\beta}_{\mathrm{LATE}}$ \\
    \hline
    LASSO & 0.0165 & 0.0209 \\
          & (0.0132) & (0.0134) \\
    Elastic Net & 0.0175 & 0.0161 \\
          & (0.0132) & (0.0146) \\
    Ridge & 0.0194 & 0.0366 \\
          & (0.0132) & (0.0133) \\
    Random Forest & 0.0215 & 0.0230 \\
          & (0.0132) & (0.0129) \\
    Regression Tree & 0.0287 & 0.0290 \\
          & (0.0134) & (0.0130) \\
    XGBoost & 0.0238 & 0.0255 \\
          & (0.0134) & (0.0133) \\
    \hline
    Short-Stacking & 0.0200 & 0.0215 \\
          & (0.0132) & (0.0128) \\
    \hline
\end{tabular}
\begin{footnotesize}
\begin{tablenotes}[flushleft]
\item \textit{Notes:} The data are \citeauthor{Abadie2003}'s (\citeyear{Abadie2003}) subsample of the 1991 Survey of Income and Program Participation (SIPP)\@. The sample size is 9,275. The outcome is an indicator for participation in IRAs, as in Table 3 of \cite{Abadie2003}. $\hat{\beta}_{\mathrm{IV}}$ and $\hat{\beta}_{\mathrm{LATE}}$ are obtained from the partially linear and interactive instrumental variables models, respectively, estimated via \texttt{ddml} with 3 cross-fitting folds. Propensity scores are trimmed at 0.01. The covariate dictionary follows the flexible specification in Table \ref{tab:401(k) reset}. Short-stacking assigns nonnegative weights to all six machine learning algorithms \citep{AhrensHansenSchafferWiemann2025}.
\end{tablenotes}
\end{footnotesize}
\end{threeparttable}
\end{adjustwidth}
\end{table}

Table \ref{tab:401(k) dml} illustrates these considerations by reporting double machine learning estimates of $\beta_{\mathrm{IV}}$ and $\beta_{\mathrm{LATE}}$ in the 401(k) application considered above, where $\beta_{\mathrm{IV}}$ denotes the variance-weighted local average treatment effect obtained with nonparametric covariate adjustment. In the case of both parameters, we estimate the relevant conditional means using six different machine learning algorithms: LASSO, elastic net, ridge, random forest, regression tree, and XGBoost. The first three methods are penalized regressions that differ in how they shrink or select coefficients, with elastic net combining the approaches of LASSO and ridge. The remaining three methods are tree-based and allow for nonlinearities and interactions without specifying their functional forms in advance, with random forest and XGBoost combining multiple trees. Thus, the differences across algorithms do not simply reflect a ranking by model flexibility.

The resulting estimates in Table \ref{tab:401(k) dml} are generally smaller than the initial estimate of $\beta_{\mathrm{IV}}$ in Panel A of Table \ref{tab:401(k) reset} but larger than the corresponding estimates of $\beta_{\mathrm{LATE}}$\@. At the same time, the double machine learning estimates remain somewhat sensitive to the choice of the machine learning algorithm. For $\beta_{\mathrm{IV}}$, the tree-based methods yield somewhat larger estimates than the penalized regressions, although this pattern does not universally hold for $\beta_{\mathrm{LATE}}$\@. Because no single machine learning algorithm performs best in all settings, we recommend that researchers routinely report estimates from at least two or three algorithms to assess the sensitivity of their conclusions.

As an alternative to reporting multiple estimates, researchers can use stacking, a model averaging approach that combines several algorithms into a single estimate. The final row of Table \ref{tab:401(k) dml} reports the short-stacking estimate recommended by \cite{AhrensHansenSchafferWiemann2025}. Rather than relying on any single algorithm, stacking lets the data determine the optimal combination of estimates and therefore provides a summary of the evidence across methods. Specifically, short-stacking assigns greater weight to algorithms that improve the predictive accuracy of the combined predictions, using cross-fitted predictions to determine these weights. The weights are constrained to be nonnegative and sum to one, which helps limit overfitting, although data-driven weighting may still overfit in finite samples. As a simple benchmark, equally averaging the six individual estimates in Table \ref{tab:401(k) dml} yields 0.0212 for $\beta_{\mathrm{IV}}$ and 0.0252 for $\beta_{\mathrm{LATE}}$, compared with 0.0200 and 0.0215 from short-stacking. The similarity suggests that our substantive conclusions are not sensitive to the weighting procedure.

\section*{Assumption Violations}

Beyond functional form, researchers should assess the identifying assumptions underlying the local average treatment effect framework. In this section, we discuss an estimation approach that accommodates some violations of monotonicity, then review testable implications of the identifying assumptions and their implementation.

\subsection*{Robustness to Monotonicity Violations}

The monotonicity assumption in instrumental variables analysis is that the instrument $Z_{i}$ must shift the treatment in the same direction for everyone who reacts to it. This implies that there are no defiers, who respond to the instrument by moving in the opposite direction. If the monotonicity assumption is rejected or otherwise implausible, estimation of average treatment effects will be difficult, and the previously discussed estimators of $\beta_{\text{LATE}}$ and $\beta_{\text{IV}}$ will not consistently estimate any ``weakly causal'' target parameter. At the same time, appropriate estimators of $\beta_{\text{AI}}$ will not be subject to such concerns, at least when a weaker assumption, termed ``weak monotonicity'' in \cite{Sloczynski2026}, is assumed to hold. Weak monotonicity allows both compliers and defiers to exist, but they must not coexist at any value of covariates; that is, weak monotonicity requires that there be no defiers at some covariate values and no compliers everywhere else. This assumption is strictly weaker than the usual (``strong'') monotonicity, which rules out the existence of defiers everywhere. Even so, weak monotonicity will be plausible in some applications, but too restrictive in others.

To see why standard estimators of $\beta_{\text{LATE}}$ and $\beta_{\text{IV}}$ will not work under weak monotonicity, note that the term $\pi_j$ in the weighted average representations in Table \ref{tab:LATE-weights} is the conditional first stage, which equals the conditional proportion of compliers minus the conditional proportion of defiers. Under strong monotonicity, there are no defiers, so $\pi_j$ is the conditional proportion of compliers and is therefore nonnegative. However, under weak monotonicity, the conditional first stage may be positive at some covariate values and negative at others, which directly translates to the incidence of ``negative weights'' in  $\beta_{\text{LATE}}$ and $\beta_{\text{IV}}$ (as evident in Table \ref{tab:LATE-weights}). On the other hand, $\pi_j$ enters the expression in $\beta_{\text{AI}}$ as a quadratic rather than linearly, which ensures that the resulting weights are nonnegative even if some conditional first stages are negative. \cite{Sloczynski2026} argues that this makes focusing on $\beta_{\text{AI}}$ a useful compromise under weak monotonicity.

This recommendation requires an additional explanation. The weighted average representations of $\beta_{\text{LATE}}$, $\beta_{\text{IV}}$, and $\beta_{\text{AI}}$ are expressed as $\beta_m = \sum_j w_{m,j} \cdot \tau_j$, where the weights $w_{m,j}$ are normalized versions of the unnormalized weights, $\omega_{m,j}$, in Table \ref{tab:LATE-weights}. Under weak monotonicity, not only does $\pi_j$ change its interpretation, but $\tau_j$ does as well. Indeed, $\tau_j$ should now be understood as the average treatment effect for either compliers or defiers in cell $j$, \emph{whichever group resides in that cell}. This also implies that $\beta_{\text{LATE}}$, defined as $\beta_{\text{LATE}} = \sum_j p_j \pi_j \tau_j / \sum_j p_j \pi_j$, should no longer be interpreted as the average treatment effect for compliers (or compliers and defiers). Instead, it corresponds to the probability limit of a generic estimator of the local average treatment effect, derived under the assumption of strong monotonicity. Such an estimator will be inconsistent if strong monotonicity is violated but weak monotonicity is maintained.

A numerical illustration helps clarify these issues. Consider a stratified randomized controlled trial with noncompliance and two equal-sized covariate cells. In both cells, 50 percent of individuals receive the instrument. In the first cell, 20 percent of individuals are compliers, there are no defiers, and the average treatment effect for compliers is 1. In the second cell, there are no compliers, 10 percent of individuals are defiers, and their average treatment effect is 4. The average treatment effect for the combined complier--defier subpopulation is 2. However, $\beta_{\mathrm{LATE}} = \beta_{\mathrm{IV}} = -2$, because these estimands attach a negative weight to the average treatment effect of the second-cell defiers. Even though both conditional average treatment effects are positive, these estimands are negative. By contrast, $\beta_{\mathrm{AI}} = 1.6$. While not exactly equal to the average treatment effect for the combined complier--defier subpopulation, this estimand provides a much better summary of the causal effects of treatment.

However, focusing on $\beta_{\text{AI}}$ also raises practical concerns. Recall that $\beta_{\text{AI}}$ is based on the standard linear specification considered earlier, with interactions of $X_i$ and $Z_i$ as instruments for $D_i$. Because these interactions can generate many effective instruments relative to the sample size, two-stage least squares estimation of $\beta_{\text{AI}}$ may overfit the first stage and suffer from many-instrument bias. In this journal, \cite{GPHK2026} explain why conventional weak-instrument diagnostics based on the first-stage $F$ statistic are not directly applicable with many instruments and recommend a particular ``jackknife'' instrumental variables estimator as an alternative to two-stage least squares. The basic idea of jackknife estimation is straightforward: if we ``leave out'' one observation at a time when estimating the first stage, we will reduce overfitting and effectively obtain a separate first stage for each observation. While there are many different methods based on the jackknife, the two most appropriate estimators in the present context are the unbiased jackknife instrumental variables estimator (UJIVE) of \cite{Kolesar2013} and the fixed effect jackknife instrumental variables (FEJIV) estimator of \cite{CSW2023}. In addition, in order to test whether the interacted instruments are jointly strong enough for consistent estimation, we can use the pretest proposed by \cite{MS2022}. Although the pretest is not formally designed for settings with many covariates (in addition to many instruments), the simulations in \cite{Sloczynski2026} suggest that it effectively distinguishes cases in which FEJIV and UJIVE perform well from those in which all estimators of $\beta_{\text{AI}}$ perform poorly.

In Appendix Table \ref{tab:manyiv}, we provide a numerical illustration of these issues using data from \cite{Stevenson2018}. We use assignment to the most lenient bail judge as an instrument to estimate the causal effects of pretrial detention on incarceration length. The standard estimate of $\beta_{\text{IV}}$ suggests that pretrial detention leads to a large, statistically significant increase in incarceration length of about 666 days. However, strong monotonicity is likely violated in this setting, as discussed by \cite{Stevenson2018} and \cite{Sloczynski2026}. When we instead focus on estimating $\beta_{\text{AI}}$, we obtain a two-stage least squares estimate of about 130 days, which is 80 percent smaller than $\hat{\beta}_{\mathrm{IV}}$. However, two-stage least squares likely suffers from many-instrument bias. When we correct for this source of bias using the two jackknife estimators, we obtain an even smaller estimated increase in incarceration length of 51--56 days, which is not statistically different from zero.

\subsection*{Testing Local Average Treatment Effect Assumptions}

As first observed by \cite{BP1997}, the standard local average treatment effect assumptions---specifically, independence of the instrument, exclusion restriction, and monotonicity---imply two inequalities that can be used to test instrument validity. These testable implications are necessary, but not sufficient, for instrument validity: that is, rejection of these implications refutes instrument validity, but a failure to reject should not be interpreted as evidence of validity, even with large data.

The two key inequalities in \cite{BP1997} compare specific joint probabilities given $Z_i=1$ and given $Z_i=0$. The first inequality requires that if the local average treatment effect assumptions hold, then the probability that an individual is treated and their outcome assumes a value in a particular set given $Z_i=1$ is greater than or equal to the same probability given $Z_i=0$. The second inequality reverses this requirement for the probability that an individual is \emph{untreated} and their outcome assumes a value in a particular set. Both inequalities must hold for every possible set of outcome values. In what follows, we will often refer to these inequalities as requiring that the difference between each pair of probabilities is nonnegative, which is equivalent to the statements above.

As an example, in the Oregon Health Insurance Experiment, we would require that the fraction of individuals who are enrolled in Medicaid and visited their doctor more than twice in the last six months is at least as large for lottery winners as for lottery losers. We would also require that the fraction of individuals who are \emph{not} enrolled in Medicaid and never visited their doctor in the last six months is at least as large for lottery losers as for lottery winners. Importantly, we would also require that both conditions hold for any other set of possible outcome values, not just ``more than two'' and ``zero.''

Why is it the case that these inequalities must hold whenever an instrument is valid? One explanation follows from the theoretical results in \cite{IR1997}, who observed that the difference between the two probabilities in the first inequality identifies the joint probability that an individual is a complier and their \emph{treated} outcome is in the set under consideration. Similarly, the difference between the two probabilities in the second inequality identifies the joint probability that an individual is a complier and their \emph{untreated} outcome is in the respective set. For example, the difference between the fractions of lottery winners and lottery losers who are enrolled in Medicaid and visited their doctor more than twice in the last six months identifies the fraction of individuals who comply with the lottery instrument and visit their doctor more than twice in six months when enrolled in Medicaid. Put simply, if the instrument is valid, the difference between two specific probabilities (which are easy to compute) must be equal to another probability (which is otherwise impossible to compute), and we know that every probability must be nonnegative. This is exactly what the \citeauthor{BP1997} conditions require.

These testable implications have been the focus of subsequent work. \cite{Kitagawa2015} derived a formal statistical test based on the \citeauthor{BP1997} conditions, and proved that they are ``sharp'' (that is,  cannot be improved upon) as well as necessary but not sufficient for instrument validity. \cite{HM2015} developed a sharp test of weaker assumptions that still suffice to identify the local average treatment effect. \cite{MW2017} reformulated the inequalities in \cite{BP1997} in a way that facilitates implementation. \cite{Sun2023} and \cite{KR2024} extended this framework to the case of multivalued treatments.

The resulting statistical tests may be difficult to implement in some applications, especially when the instrument is valid only conditional on covariates. In such cases, the covariates enter the key inequalities as conditioning variables, which increases the computational burden. To facilitate implementation, \cite{FGK2022} proposed testing the \citeauthor{BP1997} conditions using machine learning, which makes it possible to detect violations that occur only at some covariate values and would otherwise be masked by aggregation.

Statistical tests based on the \citeauthor{BP1997} conditions (and their extensions) remain underused in practice.\footnote{As discussed earlier, it is important to remember that these testable implications should be interpreted with caution. They are necessary but not sufficient for instrument validity, so a failure to reject should never be used to suggest that the instrument is valid. In addition, a failure to reject is especially uninformative in settings with strong first stages \citep[cf.][]{KR2024}. In the limiting case of perfect compliance, the testable implications are automatically satisfied.} It is slightly more common among practitioners to consider the implication of independence and (strong) monotonicity that the conditional first stage must be nonnegative at all covariate values. In the absence of monotonicity, the conditional first stage identifies the conditional difference between the fractions of compliers and defiers. However, if we assume that there are no defiers, this difference must be nonnegative everywhere. Practitioners sometimes examine whether this is the case, but the groups they consider are typically chosen ad hoc. \cite{Andresen2026} provides the R package \texttt{montest} that enables the use of machine learning to select the groups for testing; the package also implements other testable implications of the local average treatment effect assumptions, including those in \cite{BP1997}, \cite{MW2017}, and \cite{KR2024}.

In Appendix Table \ref{tab:montest}, we offer an illustration of these tests in three empirical applications, showing that there is no evidence against instrument validity in two of the applications, while the null hypothesis of instrument validity is clearly rejected in a third.

\section*{Best Practices}

In this section, we bring together the recommendations developed throughout this paper in five practical guidelines for instrumental variables applications with heterogeneous treatment effects. Table \ref{tab:packages} lists selected software packages for implementing these recommendations.

\paragraph{Don't automatically call an instrumental variables estimate ``the LATE.''} With covariates, standard instrumental variables regressions generally recover weighted averages that differ from the average effect for all compliers. When the standard LATE assumptions are credible, we recommend targeting this latter parameter explicitly, or at least estimating it as a robustness check. Estimate the average effect of the instrument on the outcome and divide by its average effect on treatment. With saturated covariates, our LATE specification implements this approach in a single regression.

\paragraph{Check functional forms, even when the controls are right.} Including the relevant covariates does not ensure that they enter the model appropriately. Apply RESET to the instrument propensity score specification, using its nonlinear version for logit or probit, and consider richer specifications when it rejects. Covariate balancing and doubly robust estimators offer additional protection against misspecification.

\paragraph{Use machine learning, but check sensitivity to the algorithm.} Double machine learning offers a flexible alternative to manually choosing interactions and higher-order terms. Use orthogonal scores and cross-fitting, and report estimates from at least two or three algorithms. Alternatively, use stacking to combine their predictions. Flexibility does not eliminate dependence on approximation assumptions.

\paragraph{Reconsider the estimator when monotonicity is doubtful.} If the instrument encourages treatment in some groups but discourages it in others, conventional LATE estimators can attach negative weights to causal effects. When compliers and defiers do not coexist at any covariate value, \citeauthor{AI1995}'s specification with saturated covariates instead recovers a nonnegatively weighted average of their effects. This is a useful compromise, although the target parameter changes. When interactions generate many instruments, use appropriate jackknife estimators (UJIVE or FEJIV) and many-instrument diagnostics.

\paragraph{Test the identifying assumptions, but do not treat nonrejection as validation.} Complement substantive arguments with formal tests of the LATE assumptions, including the \citeauthor{BP1997} inequalities and restrictions on conditional first stages. Account for covariates when validity is conditional; machine learning can help detect violations confined to particular groups. These tests can uncover contradictions with the assumptions, but cannot establish instrument validity.

\section*{Conclusion}

The local average treatment effect framework has changed how researchers interpret instrumental variables estimates. Realizing its promise also requires changes in empirical practice: choosing estimators that target substantively meaningful effects, allowing flexible covariate adjustment, and confronting the identifying assumptions with the data. These steps can materially change the conclusions drawn from familiar research designs.

Interested readers may also wish to explore other survey papers on instrumental variables methods. For example, \cite{MT2024} review the literature on instrumental variables methods with heterogeneous treatment effects, including extensions to multivalued treatments, multivalued instruments, and multiple instruments. \cite{MT2018} review the ``marginal treatment effect'' framework of \cite{HV2005}; the marginal treatment effect is especially relevant with continuous instruments and can be understood as the LATE for a small change in the instrument. In this journal, \cite{BHJ2025} provide a guide to the literature on shift-share instrumental variables. \cite{CFL2025} and \cite{GPHK2026} survey the literature on leniency designs, another important instrumental variables application involving quasi-random assignment to many decision-makers rather than a single binary instrument.

\section*{Acknowledgments}

We thank Martin Andresen, Kirill Borusyak, Peter Hull, Jeffrey Kling, Jonathan Parker, Jonathan Roth, Timothy Taylor, Alex Torgovitsky, Heidi Williams, and Jeff Wooldridge for helpful feedback. Liyang Sun acknowledges generous support from the James M. and Cathleen D. Stone Centre on Wealth Concentration, Inequality, and the Economy at University College London through an Internal Research Support Grant (IRSG/2526/LIY)\@. Derya Uysal acknowledges generous support from Deutsche Forschungsgemeinschaft (CRC TRR 190, project number 280092119).

\begin{landscape}
\thispagestyle{empty}
\begin{table}[p]
\centering
\small
\makebox[\linewidth][c]{%
\begin{threeparttable}
\caption{Selected Software Packages in MATLAB, R, and Stata
\label{tab:packages}}
\begin{tabular}{
    >{\centering\arraybackslash}m{3cm}
    >{\centering\arraybackslash}m{2.5cm}
    >{\centering\arraybackslash}m{8.5cm}
    >{\centering\arraybackslash}m{10cm}
}
    \hline\hline
    \textbf{Package Name} &
    \textbf{Software} &
    \textbf{Description} &
    \textbf{Download Information} \\
    \hline
    \\
    \multicolumn{4}{l}{A. Estimating the LATE} \\
    \hline
    \texttt{causalweight} & R & IPW estimation of the LATE using function \texttt{lateweight} & CRAN \\
    \hline
    \texttt{drlate} & Stata & Regression adjustment, IPW, and doubly robust estimation of the LATE & SSC \\
    \hline
    \texttt{kappalate} & Stata & IPW estimation of the LATE & SSC \\
    \hline
    \texttt{lateffects} & Stata & IPW and doubly robust estimation of the LATE & StataNow 19 \\
    \hline
    \\
    \multicolumn{4}{l}{B. Double Machine Learning} \\
    \hline
    \texttt{ddml} & R and Stata & Double machine learning estimation of IV and LATE parameters & CRAN (R); SSC (Stata) \\
    \hline
    \texttt{DoubleML} & R & Double machine learning estimation of IV and LATE parameters & CRAN \\
    \hline
    \\
    \multicolumn{4}{l}{C. Testing LATE Assumptions} \\
    \hline
    \texttt{LATEtest} & R & Tests of the LATE assumptions based on machine learning & \url{https://github.com/farbmacher/LATEtest} \\
    \hline
    \texttt{montest} & R & Tests of monotonicity and LATE assumptions based on machine learning & \url{https://github.com/martin-andresen/montest} \\
    \hline
    \texttt{TestMechs} & R & Tests of mediation mechanisms and related testable implications of LATE assumptions & \url{https://github.com/jonathandroth/TestMechs} \\
    \hline
    \\
    \multicolumn{4}{l}{D. Estimation with Many Instruments} \\
    \hline
    \texttt{fejiv} & MATLAB, R, and Stata & Fixed-effect jackknife IV estimation with many instruments & MATLAB Central File Exchange; CRAN (R); SSC (Stata) \\
    \hline
    \texttt{manyiv} & Stata & IV estimation and inference with many instruments & \url{https://github.com/gphk-metrics/stata-manyiv} \\
    \hline
    \texttt{manyweakiv} & Stata & Pretest for weak identification with many instruments using command \texttt{manyweakivpretest} & \url{https://github.com/lsun20/manyweakiv} \\
    \hline
\end{tabular}
\begin{tablenotes}[flushleft]
\item \textit{Notes:} ``CRAN'' refers to the Comprehensive R Archive Network. To download an R package from CRAN, type \texttt{install.packages("package")} in the R/RStudio console, where \texttt{package} is a placeholder for the package name (e.g., \texttt{causalweight}). ``SSC'' refers to the Statistical Software Components Archive. To download a Stata package from SSC, type \texttt{ssc install package} in the Command window, where \texttt{package} is a placeholder for the package name (e.g., \texttt{drlate}). ``StataNow 19'' refers to the continuous-release version of Stata 19, which added the \texttt{lateffects} command in August 2025.
\end{tablenotes}
\end{threeparttable}%
}
\end{table}
\end{landscape}

\singlespacing

\setlength\bibsep{0pt}
\bibliographystyle{aer}
\nocite{Abadie_data,OHIE_data}
\bibliography{LATE_references}

\newpage

\onehalfspacing

\setcounter{table}{0}
\renewcommand{\thetable}{A\arabic{table}}
\renewcommand{\tablename}{Appendix Table}

\begin{table}[!p]
\begin{adjustwidth}{-1in}{-1in}
\centering
\begin{threeparttable}
\caption{Causal Effects of Pretrial Detention on Incarceration Length \normalsize{\label{tab:manyiv}}}
\begin{tabular}{>{\centering\arraybackslash}m{2.5cm} >{\centering\arraybackslash}m{2.5cm} >{\centering\arraybackslash}m{2.5cm} >{\centering\arraybackslash}m{2.5cm}}
    \hline\hline
    \multirow{2}[2]{*}{$\hat{\beta}_{\mathrm{IV}}$} & \multicolumn{3}{c}{$\hat{\beta}_{\mathrm{AI}}$} \\
    \cline{2-4}
          & 2SLS & FEJIV & UJIVE \\
    \hline
	666 & 130 & 51 & 56 \\
	(233) & (43) & (91) & (99) \\
	\hline
\end{tabular}
\begin{footnotesize}
\begin{tablenotes}[flushleft]
\item \textit{Notes:} The data are \citeauthor{Stevenson2018}'s (\citeyear{Stevenson2018}) sample of 331,971 arrests in Philadelphia. The outcome is incarceration length, defined as the maximum days of an incarceration sentence. The treatment is pretrial detention. The instrument is whether a given case was heard by the most lenient judge, referred to as ``Judge C'' in \cite{Sloczynski2026}. The covariate specification is saturated in the 17 most common offense types, race and gender of the defendant, and three time periods considered by \cite{Stevenson2018}. Groups with fewer than three cases heard by ``Judge C'' or not heard by ``Judge C'' are dropped. 2SLS denotes two-stage least squares; FEJIV and UJIVE denote the fixed effect jackknife instrumental variables estimator of \cite{CSW2023} and the unbiased jackknife instrumental variables estimator of \cite{Kolesar2013}, respectively. Estimates and standard errors (in parentheses) are reproduced from specification \#3 in Panel B of Table 8 in \cite{Sloczynski2026}.
\end{tablenotes}
\end{footnotesize}
\end{threeparttable}
\end{adjustwidth}
\end{table}

\begin{table}[!p]
\begin{adjustwidth}{-1in}{-1in}
\centering
\begin{threeparttable}
\caption{$p$-Values for Tests of Instrument Validity \normalsize{\label{tab:montest}}}
\begin{tabular}{>{\centering\arraybackslash}m{5.5cm} >{\centering\arraybackslash}m{3cm} >{\centering\arraybackslash}m{3cm} >{\centering\arraybackslash}m{3cm}}
    \hline\hline
     & BP & MW & FS \\
    \hline
	\cite{Finkelsteinetal2012} & 0.99996 & 0.971 & 0.9999998 \\
    \cite{Abadie2003} & 0.997 & 0.9999996 & N/A \\
    \cite{Stevenson2018} & 0.000102 & 0.0241 & 0.000226 \\
	\hline
\end{tabular}
\begin{footnotesize}
\begin{tablenotes}[flushleft]
\item \textit{Notes:} The source of data and variable choice for \cite{Finkelsteinetal2012} are the same as in Panel A of Table \ref{tab:OHIE main}. The source of data and variable choice for \cite{Abadie2003} are the same as in Tables \ref{tab:401(k) reset} and \ref{tab:401(k) dml}. For \cite{Stevenson2018}, the data are a sample of 331,971 arrests in Philadelphia. The outcome is incarceration length, defined as the maximum days of an incarceration sentence. The treatment is pretrial detention. The instrument is whether a given case was heard by the most lenient judge, referred to as ``Judge C'' in \cite{Sloczynski2026}. The covariate specification is saturated in the 17 most common offense types, race and gender of the defendant, and three time periods considered by \cite{Stevenson2018}. Groups with fewer than three cases heard by ``Judge C'' or not heard by ``Judge C'' are dropped. ``BP'' refers to the \citeauthor{BP1997} conditions. ``MW'' refers to the \citeauthor{MW2017} conditions. ``FS'' refers to the requirement that the conditional first stage is nonnegative at all covariate values. All tests are implemented using the \texttt{montest} command in R \citep{Andresen2026}.
\end{tablenotes}
\end{footnotesize}
\end{threeparttable}
\end{adjustwidth}
\end{table}

\end{document}